\documentclass[prd,twocolumn,nofootinbib]{revtex4}
\usepackage{graphicx, epsfig}
\usepackage{color}
\usepackage{mathrsfs}
\usepackage{bm}
\usepackage{amsmath,amssymb,yhmath}
\usepackage[caption=false]{subfig}
\usepackage{hyperref}
\usepackage[normalem]{ulem}
\usepackage{mathtools}

\newcommand{\be}{\begin{equation}}
\newcommand{\ee}{\end{equation}}
\newcommand{\ba}{\begin{eqnarray}}
\newcommand{\ea}{\end{eqnarray}}

\newcommand{\half}{\frac{1}{2}}

\begin{document}
\title{Rolling classical scalar field in a linear potential coupled to a quantum field}
%\title{Rolling with quantum fields}
\author{Mainak Mukhopadhyay, Tanmay Vachaspati}
\affiliation{
Physics Department, Arizona State University, Tempe, AZ 85287, USA.
}

\begin{abstract}
\noindent
We study the dynamics of a classical scalar field that rolls down a linear potential
as it interacts bi-quadratically with a quantum field. We explicitly solve the dynamical 
problem by using the classical-quantum correspondence (CQC). Rolling solutions on 
the effective potential are shown to compare very poorly with the full solution. Spatially 
homogeneous initial conditions maintain their homogeneity and small 
inhomogeneities in the initial conditions do not grow.
\end{abstract}

\maketitle

\section{Introduction}
\label{intro}

Often we are interested in the dynamics of quantum fields in classical backgrounds.
A prime example is that of phase transitions in which an order parameter evolves to 
develop a vacuum expectation value while also interacting with other quantum degrees 
of freedom. Another example is that of inflationary cosmology where the inflaton rolls 
down some potential while exciting other quantum fields to reheat the universe. 
Such problems have a rich history but the attention has mostly focused on the
quantum effects in fixed classical backgrounds, while it is of interest to also
examine the quantum backreaction on the evolution of the classical
background. 

%Progress has recently been made by precisely mapping the quantum
%degrees of freedom to corresponding classical degrees of freedom and in this
%way transforming the quantum plus classical problem to a fully classical problem
%\cite{Vachaspati:2018llo,Vachaspati:2018hcu}.

Progress on this mixed classical and quantum problem is possible by mapping
the quantum degrees of freedom to corresponding classical degrees of freedom,
thus obtaining a fully classical 
problem~\cite{Aarts:1999zn,Vachaspati:2018llo,Vachaspati:2018hcu}.
The solution of the full classical problem contains all information about the quantum
variables and also the backreacted dynamics of the classical background. The
method has been illustrated in a few applications so far: 
backreaction of fermion production on gauge fields~\cite{Aarts:1999zn},
quantum mechanical rolling~\cite{Vachaspati:2018llo} where the method
was explicitly tested, Hawking evaporation during gravitational
collapse~\cite{Vachaspati:2018pps}, and the quantum evaporation of field theory 
defects~\cite{Borsanyi:2007wm} and oscillons~\cite{Hertzberg:2010yz,Olle:2019skb}. 
Here we will consider rolling in field theory;
some other works on this problem using different approaches and approximations 
can be found in Refs.~\cite{Bardeen:1986iq,Boyanovsky:1992vi,Mrowczynski:1994nf,
Ramsey:1997sa,Bedingham:2002qn,Aarts:2007ye,Asnin:2009bs}. 

%The technique to map the quantum problem to a classical problem, also
%known as the classical-quantum correspondence (CQC), is similar to the
%semiclassical approach in which the classical background dynamics couples 
%to the expectation value of the quantum operators in the equation of motion.
The technique to map the quantum problem to a classical problem, 
which we describe as a classical-quantum correspondence (CQC)
can be done by using mode functions~\cite{Aarts:1999zn}, or equivalently,
by going to classical variables in higher 
dimensions~\cite{Vachaspati:2018llo,Vachaspati:2018hcu}. The background
dynamics is assumed to be well described by the semiclassical
approximation in which the classical background dynamics couples 
to the expectation value of the quantum operators in the equation of 
motion. The expectation value is also evaluated in the dynamical
background in terms of the classical variables.
The validity of this approach has been explicitly tested in a quantum
mechanical setting where the full quantum solution can be compared to
the CQC result~\cite{Vachaspati:2018llo}.

To be more specific, we will consider a model with two scalar fields, $\phi$
and $\psi$, where $\phi$ is the classical background and $\psi$ is the
quantum field interacting with this background.
%\section{Quantum field and CQC}
%\label{qmfieldcqc}
The Lagrangian (in 1+1 dimensions) is,
\be
L = \half (\partial_\mu\phi)^2 + \half (\partial_\mu\psi)^2  - V(\phi )
-\frac{m^2}{2}\psi^2 - \half\lambda \phi^2\psi^2
\label{model}
\ee
where we will mostly focus on the case of a linear potential $V(\phi ) = - \kappa \phi$
on which $\phi$ can roll. 
One approach to solving for the dynamics is to realize that the Lagrangian is
quadratic in the quantum field $\psi$. Thus it can be integrated out in the
path integral. This will yield a term in the effective action that has the form
$\ln ( {\rm Det}{\hat O}[\phi ])$ where ${\hat O}[\phi]$ is an operator that
depends on the background $\phi$ (see~\cite{Peskin:1995ev} for example).
Usually, at this stage, one adopts a
perturbative approach and assumes $\phi$ is a known background to lowest
order in some coupling. Then it may be possible to diagonalize ${\hat O}$ and 
to evaluate $\ln ( {\rm Det}{\hat O}[\phi])$ perturbatively or in some other
approximation scheme~\cite{Bardeen:1986iq,Boyanovsky:1992vi,Mrowczynski:1994nf,
Ramsey:1997sa,Bedingham:2002qn,Aarts:2007ye,Asnin:2009bs}. 

In contrast, in the CQC, one does not try to eliminate $\psi$ from the action. 
Instead the CQC equations {\it simultaneously} evolve the background $\phi$ 
as coupled to the expectation value of $\psi^2$ and the quantum 
operator $\psi$ in the $\phi$ background.
This becomes possible by rewriting the quantum operator $\psi$ in terms of new
c-number variables denoted by a complex matrix $Z$ and the initial quantum
operators. The evolution of $\psi$ is 
given entirely by the evolution of $Z$~\cite{Vachaspati:2018hcu} and the
expectation of $\psi^2$ that enters the $\phi$ equation takes the form $Z^*Z$.
%
%Instead one replaces the quantum term $\psi^2$ in the equation of motion for
%$\phi$ by its expectation value $\langle \psi^2 \rangle$ which is then written in 
%terms of new variables denoted by a complex matrix $Z$. 
%
%One can show that the $Z$ variables
%satisfy certain classical equations of motion that themselves contain the 
%unknown background $\phi$~\cite{Vachaspati:2018hcu}. 
%
In this way, we obtain a set of differential equations 
for $\phi$ and $Z$ that are solved with specific initial conditions to obtain the full 
dynamics. The background is completely general and need not be homogeneous,
and perturbation theory is not employed. The only assumption is that the 
background is classical and it couples to the expectation value of $\psi^2$
evaluated in its dynamical quantum state.

To understand the CQC equations more quantitatively, we write the semiclassical 
equation of motion
for $\phi$,
\be
\square \phi + V'(\phi) + \lambda \langle \psi^2 \rangle \phi =0,
\label{phieq}
\ee
where the expectation value $\langle \psi^2 \rangle$ is in the (unknown) instantaneous
quantum state for the $\psi$ fields\footnote{The quantity $\langle \psi^2 \rangle$ is
formally divergent but the divergence can be absorbed by mass renormalization. We 
will discuss renormalization in Sec.~\ref{renormalization}.}. 
The evolution of the quantum operator $\psi$ is
given by the Heisenberg equation,
\ba
{\dot \psi} &=& \pi, \\
{\dot \pi } &=& \nabla^2 \psi + (m^2 + \lambda \phi^2)\psi
\label{Heqs}
\ea
where $\pi$ denotes the conjugate momentum to $\psi$.
This equation for quantum $\psi$ can be solved in terms of a 
c-number variable in two spatial dimensions, $Z(t,x,y)$, by 
writing\footnote{If one Fourier transforms on the $y$ variable,
$Z$ will map on to the usual mode functions~\cite{Aarts:1999zn}. 
This can be useful in homogeneous backgrounds. We find that
the discretization discussed in Sec.~\ref{CQCdiscrete} is more 
intuitive while thinking of $y$ as a spatial coordinate.},
\be
\psi(t,x) = \int dy\, \left [ Z^*(t,x,y) a_0(y) + Z(t,x,y) a_0^\dag (y) \right ]
\label{psiZ}
\ee
where $a_0$ and $a_0^\dag$ are annihilation and creation operators
at the initial time. $a_0$ is defined by
\be
a_0(y) = \frac{1}{\sqrt{2}} \left ( \sqrt{\Omega_0}^{\,-1} \, \pi_0 (y) 
-i \sqrt{\Omega_0}\,  \psi_0(y) \right )
\ee
and $a_0^\dag(y)$ is the Hermitian conjugate of $a_0$. Also
$\Omega_0^2 = \nabla_y^2 + m^2 + \lambda \phi_0^2$ and $\phi_0 = \phi (t=0,y)$
may depend non-trivially on $y$. Inserting \eqref{psiZ} in \eqref{Heqs} we obtain
the equation of motion for $Z$,
\be
{\ddot Z} - \nabla_x^2 Z + ( m^2 + \lambda \phi^2 ) Z = 0
\label{eqZ}
\ee
Thus $Z$ satisfies the classical equation of motion (independently of $y$).
As discussed in Sec. 4 of Ref.~\cite{Vachaspati:2018hcu} there are constraints 
on the $Z$ that arise from the field commutation relations but these are
consistent with the dynamical equations -- if the constraints are satisfied
initially, they remain satisfied on evolution.
The initial conditions for $Z$ can be obtained from the initial conditions
for $\psi$ and $\pi$, and we will write these explicitly in Sec.~\ref{CQCdiscrete}.

Next we assume that
the initial state is the vacuum and is annihilated by $a_0$. Then we find
\be
\langle \psi^2 \rangle = \int dy \, Z^* Z
\ee
where recall that we are working in the Heisenberg picture so the
quantum state at all times is given by the initial vacuum state.
Inserting the expectation value in \eqref{phieq} gives,
\be
\square \phi + V'(\phi) + \lambda \left ( \int dy \, Z^* Z \right ) \phi =0
\label{phieqZ}
\ee
So the CQC equations consist of \eqref{phieqZ} and \eqref{eqZ}. In practice
these need to be solved numerically for which they must be discretized. A
convenient discretization is discussed in Sec.~\ref{CQCdiscrete}.

%
%However, as shown in~\cite{Vachaspati:2018hcu},
%we can rewrite the quantum operator $\psi$ in terms of a complex valued matrix, $Z$,
%and quantum operators at the initial time. The matrix $Z$ can be shown to satisfy
%Eq.~\eqref{zcqceq}. In terms of $Z$ we can evaluate $\langle \psi^2 \rangle$ to get,
%\be
%\langle \psi^2 \rangle =  \frac{1}{a^2} \sum_{j=1}^N Z_{nj}^* Z_{nj} 
%\ee
%Hence \eqref{phieq} discretizes to \eqref{phicqceq} and $Z$ satisfes
%\eqref{zcqceq} with the initial conditions in \eqref{Zinitial}, thus giving
%us a closed system of differential equations that we need to solve.

A simplification occurs if attention is restricted to static solutions. Then the
background is fixed and the CQC approach is equivalent to the effective 
potential.
It is only when we are interested in dynamical questions that the
CQC becomes a powerful tool. For example, if we consider the model
in Eq.~\eqref{model}, we can find static solutions for $\phi$ by locating the
extrema of the effective potential, or equivalently by finding static solutions
to the CQC equations. If on the other hand, we want to know
the dynamical solution, the effective potential 
is not useful whereas the CQC approach leads to the solution. The 
underlying reason is that the effective potential assumes the 
quantum state of the fields, for example the vacuum state or a thermal
state, and expectation values of operators are taken in this state.
In a dynamical process, the quantum state itself will be determined by
the dynamics and will in general be different from the vacuum (or other)
state assumed in the calculation of the effective potential. One situation
where the effective potential may suffice is if there is dissipation in the
system (for example, an expanding universe) and then the quantum
fields are consistently driven to their vacuum state. Even in this case, the
CQC can be used to describe the approach to the asymptotic state
whereas the effective potential can only describe the final asymptotic 
state after the quantum fields have dissipated into their vacuum state.

In this paper, we start by describing the discretized CQC formulation in 
Sec.~\ref{CQCdiscrete}. Then we discuss static solutions in Sec.~\ref{statics}.
This exercise is completely equivalent to the effective potential
formulation. In Sec.~\ref{dynamics} we first discuss homogeneous dynamics. 
This leads to a very different picture from that obtained by simply
considering static solutions of the effective potential. In Sec.~\ref{dynamics}
we also study dynamics with inhomogeneous initial conditions to see
if homogeneous solutions might be unstable to developing inhomgeneities.
We do not find an instability and this means that translational invariance
is not spontaneously broken. We conclude in Sec.~\ref{conclusions}.

\section{Lattice CQC}
\label{CQCdiscrete}

%We will treat $\phi$ as a classical background field and $\psi$ as a quantum field. Then
%we can apply the CQC. 
%As in Sec.~\ref{intro}, we expect field levitation to occur at strong coupling.

The CQC reformulation of the system in \eqref{model} follows that 
in~\cite{Vachaspati:2018hcu,Olle:2019skb}. One difference is that we will
employ periodic boundary conditions whereas Dirichlet boundary
conditions were used in Refs.~\cite{Vachaspati:2018hcu,Olle:2019skb}.

The first step is to latticize the field theory. The lattice points are given by $x=n a$ 
where $n=1,\ldots,N$. The discrete Lagrangian is
\ba
L' = && a \sum_{n=1}^N \biggl [ \half {\dot \phi}_n^2 
+ \frac{1}{2a^2} \phi_n \left ( \phi_{n+1} - 2\phi_n + \phi_{n-1} \right )
\nonumber \\ && \hskip 0.5 cm
+ \half {\dot \psi}_n^2 
+ \frac{1}{2a^2} \psi_n \left ( \psi_{n+1} - 2\psi_n + \psi_{n-1} \right )
\nonumber \\ && \hskip 0.5 cm
%+ K \phi_n 
- V(\phi_n) - \frac{m^2}{2} \psi_n^2 - \frac{\lambda}{2} \phi_n^2 \psi_n^2 \biggr ]
\ea
where $V(\phi)$ is a potential for $\phi$ that we will choose later.
We assume periodic boundary conditions and $n$ should be considered
to be an integer mod $N$.

%which give
%$\phi_{n+N} = \phi_n$, $\psi_{n+N}=\psi_n$ for any $n$. 
%\TV{this gives $\phi_N=\phi_1$. Did we want $\phi_{N+1}=\phi_1$?}
%We
%have written $\lambda$ to denote that this is the bare coupling
%constant. 

The $\psi$ dependent part can be written as
\be
L_\psi ' =  a \left [
\half {\dot \Psi}^T {\dot \Psi} - \frac{1}{2} \Psi^T \Omega^2 \Psi \right ]
\ee
where $\Psi$ denotes a column vector with components $\psi_i$ and
\be
\Omega_{ij}^2 = 
\begin{cases}
+2/a^2 + m^2 + \lambda \phi_i^2, &  i=j \\
-1/a^2, & i= j\pm 1, \\ 
-1/a^2 & {i=1,j=N}; \ {i=N, j=1} \\
0, & {\rm otherwise}
\end{cases}
\label{Omsquared0}
\ee

%The initial conditions are given in terms of the square root of the matrix
%$\Omega$, for which $\Omega^2$ needs to be diagonalized (at the
%initial time). In general, this will have to be done numerically.

Using the CQC, the quantum field variables $\{\psi_i\}$ map into
$N\times N$ complex classical field variables $\{ Z_{ij} \}$
that satisfy the equation of motion~\cite{Vachaspati:2018hcu},
\be
{\ddot Z}_{nj} + \sum_{k=1}^N \Omega^2_{nk} Z_{kj} = 0.
\label{zcqceq}
\ee
The CQC equation of motion for $\phi$ is
\ba
{\ddot \phi}_n - \frac{1}{a^2} (\phi_{n+1} - 2\phi_n + \phi_{n-1}) 
+ V'(\phi_n) && \nonumber \\
%- \frac{\lambda}{2\pi} \ln ( \mu a ) \phi_n
+ \lambda \left ( \frac{1}{a^2} \sum_{j=1}^N Z_{nj}^* Z_{nj} \right ) \phi_n &=& 0.
\label{phicqceq}
\ea
%
%where $\mu$ is the renormalization energy scale such that if $a=1/\mu$ then
%the potential of $\phi$ is given by $V(\phi )$. In practice we will choose $a=1/ \mu$
%for some small $a$, so that the $\ln$ term will not contribute. 

These equations of motion have to be solved with initial conditions for $Z$
that correspond to $\psi$ being in its vacuum state,
\be
Z_0 = -i \sqrt{\frac{a}{2}} \sqrt{\Omega_0}^{-1} , \ \ 
{\dot Z}_0 = \sqrt{\frac{a}{2}} \sqrt{\Omega_0} 
\label{Zinitial}
\ee
The initial conditions for $\phi$ are fixed by the problem of interest,
\be
\phi_n = \phi_n(0), \ \ {\dot \phi}_n = {\dot\phi}_n (0).
\label{phiinitial}
\ee

The sum over $Z$'s in the last term of \eqref{phicqceq} will lead to renormalization
of the mass of $\phi$ as we will discuss in Sec.~\ref{renormalization}.

%To understand the CQC equations, 
%we write the semiclassical equation of motion for $\phi$,
%\be
%\square \phi + V'(\phi) + \lambda \langle \psi^2 \rangle \phi =0,
%\label{phieq}
%\ee
%where the expectation value $\langle \psi^2 \rangle$ is in the (unknown) instantaneous
%quantum state for the $\psi$ fields. However, as shown in~\cite{Vachaspati:2018hcu},
%we can rewrite the quantum operator $\psi$ in terms of a complex valued matrix, $Z$,
%and quantum operators at the initial time. The matrix $Z$ can be shown to satisfy
%Eq.~\eqref{zcqceq}. In terms of $Z$ we can evaluate $\langle \psi^2 \rangle$ to get,
%\be
%\langle \psi^2 \rangle =  \frac{1}{a^2} \sum_{j=1}^N Z_{nj}^* Z_{nj} 
%\ee
%Hence \eqref{phieq} discretizes to \eqref{phicqceq} and $Z$ satisfes
%\eqref{zcqceq} with the initial conditions in \eqref{Zinitial}, thus giving
%us a closed system of differential equations that we need to solve.

\section{Statics}
\label{statics}

We look for static solutions of $\phi$ but $Z_{ij}$ may be time dependent.
Then we
set ${\ddot \phi}_n$ in \eqref{phicqceq} to zero. The equation is consistent only if
we can show that the $Z-$dependent factor in the last term is time independent.
This factor is proportional to $Z Z^\dag$ and hence we define,
\be
F = Z Z^\dag
\ee
Then
\ba
{\dot F} &=& {\dot Z} Z^\dag + Z {\dot Z}^\dag, \\ 
{\ddot F} &=& 2 {\dot Z}{\dot Z}^\dag - (\Omega^2 F + F\Omega^2 ),
\ea
From the initial conditions in \eqref{Zinitial} we get
\ba
Z_0 Z_0^\dag &=& \frac{a}{2} \Omega_0^{-1}, \ \ 
{\dot Z}_0 Z_0^\dag = i\frac{a}{2} = - Z_0 {\dot Z}_0^\dag \\
{\dot Z}_0 {\dot Z}_0^\dag &=& \frac{a}{2} \Omega_0 
= \Omega_0^2 Z_0 Z_0^\dag =\Omega_0^2 F(0)
\ea
From here it is straightforward to check that ${\dot F}(0)=0={\ddot F}(0)$. Also
note that $\Omega_0^2$ and $F_0 = a \Omega_0^{-1}/2$ commute. Then all higher 
derivatives of $F$ when evaluated at the initial time will also vanish. For example,
\be
{\dddot F}_0 = -2 (\Omega_0^2 Z_0 {\dot Z}_0^\dag + {\dot Z}_0 Z_0^\dag \Omega_0^2)
- (\Omega_0^2 {\dot F}_0 + {\dot F}_0 \Omega_0^2 ) = 0
\ee
Hence it is consistent to set ${\ddot \phi}=0$ in \eqref{phicqceq} and to obtain the
static equation,
\be
- \frac{1}{a^2} (\phi_{n+1} - 2\phi_n + \phi_{n-1}) + V'(\phi_n)
+ \frac{\lambda}{2a} \Omega_{0,nn}^{-1} \phi_n = 0 
%+ \lambda \left ( \frac{1}{a^2} \sum_{j=1}^N Z_{nj}^* Z_{nj} \right ) \phi_n &=& 0 
\label{phicqceqstatic}
\ee
where there is no sum over the repeated index $n$. 

Note that $\Omega_{0,nn}$ depends on $\{\phi_i\}$. So \eqref{phicqceqstatic} is
a highly non-linear (and implicit) equation for $\phi_n$. We now discuss the solution
under the assumption that $\phi$ is homogeneous~\footnote{Inhomogeneous
solutions would also be of interest as they would represent solitons that are
supported by the quantum vacuum~\cite{Huang:1980bz}.}.

\subsection{Static homogeneous solution}
\label{hsoln}

%If $\phi$ is homogeneous, the CQC is completely equivalent to the standard
%treatment of the effective potential (e.g. Peskin \& Schroeder). In the inhomogeneous
%case too, the two approaches are equivalent. The CQC is especially useful in a 
%dynamical setting.

Under the assumption that $\phi_n$ is independent of $n$, we will 
write $\phi_n=\phi_0$. This will be a self-consistent assumption only
if $\Omega_{0,nn}^{-1}$ in \eqref{phicqceqstatic} is independent of $n$.
We now check this.

With the assumption of homogeneity, $\Omega^2$ can be diagonalized explicitly. 
For $N \ge 3$ we can write
\be
\Omega^2 = O^\dag D O
\label{om2}
\ee
where
\be
O_{lk} = \frac{1}{\sqrt{N}}e^{i lk 2\pi/N}
\label{Osol}
\ee
and
\be
D_{lk}= \left [  
\frac{4}{a^2} \sin^2\left ( \frac{\pi l}{N} \right ) + m^2+ \lambda \phi_0^2 \right ] \delta_{lk}
\label{Dsol}
\ee
Then
\be
\Omega_0^{-1} = O^\dag \sqrt{D}^{-1} O
\ee
but since $| O_{lk} |^2=1/N$ for every $l,k$ we find
%\ba
%\frac{1}{2a} \Omega_{0,nn}^{-1} &=& \frac{1}{2aN} \sum_{k=1}^N (\sqrt{D}^{-1})_{kk} \\
%&\to&
%\frac{1}{2\pi } \int_0^{\pi/2} \frac{dk}{\sqrt{b^2+\sin^2k}}
%\label{Omint}
%\ea
\ba
\frac{1}{2a} \Omega_{0,nn}^{-1} &=& 
\frac{1}{2aN} \sum_{k=1}^N (\sqrt{D}^{-1})_{kk} \\
&=&
\frac{1}{2aN} \sum_{k=1}^N 
\frac{1}{\sqrt{\frac{4}{a^2} \sin^2\left ( \frac{\pi k}{N} \right ) + M^2}}
\label{Omsum}
\ea
where 
\be
M = \sqrt{ m^2 + \lambda \phi_0^2 }
\label{Mdef}
\ee
Note that $\Omega_{0,nn}^{-1}/2a$ (no sum over $n$) is independent of $n$
when $\phi_0$ is homogeneous and the homogeneity assumption is self-consistent. 
This completes our check.

To connect with the continuum calculation we take the $a\to 0$ and
$L=aN \to \infty$ limit.
In the limit $a \to 0$, only terms with $\sin^2(\pi k/N) \to 0$ will contribute
to the sum in \eqref{Omsum}. So we can approximate $\sin^2(\pi k/N) \sim (\pi k/N)^2$.
Define $q = 2\pi k/(aN)$ and also consider the $L = aN \to \infty$ 
limit. Then
\be
\frac{1}{2a} \Omega_{0,nn}^{-1} \to
\frac{1}{2\pi} \int_0^\infty \frac{dq}{\sqrt{q^2+ M^2}}
\label{Omint}
\ee

The quantity $\Omega_{0,nn}^{-1}/2a$ in \eqref{phicqceqstatic} is completely equivalent
to the vacuum expectation value of $\psi^2$.
In the usual quantum field theory treatment,
with constant $\phi=\phi_0$, $\psi$ is a free field with mass $M$.
The standard treatment then gives
\be
\langle \psi^2 \rangle = \frac{1}{2\pi} \int_0^\infty \frac{dp}{\sqrt{p^2+M^2}}
%= \frac{1}{4\pi} \ln \left [ \frac{\sqrt{p^2+M^2}+p}{\sqrt{p^2+M^2}-p} \right ]
\label{psi2expec}
\ee
exactly as in \eqref{Omint}. 
The integral in \eqref{psi2expec} is log-divergent but the divergence can
be absorbed by mass renormalization as we now discuss.

\subsection{Renormalization}
\label{renormalization}

Eq.~\eqref{phicqceqstatic} depends on $\Omega_{0,nn}^{-1}/2a$
which is given by \eqref{Omsum}. Let us evaluate this term in the $N\to \infty$
limit while keeping the lattice spacing, $a$, fixed. Then,
\be
\frac{1}{2a} \Omega_{0,nn}^{-1} =
\frac{1}{L} \sum_{k=1}^{N/2} 
\frac{1}{\sqrt{\frac{4}{a^2} \sin^2\left ( \frac{\pi k}{N} \right ) + M^2}}
\label{Omexp}
\ee

For $\pi k/N < \pi/4$, we approximate $\sin^2(\pi k/N) \sim (\pi k/N)^2$,
while for $\pi k/N > \pi/4$ we approximate $\sin^2(\pi k/N) \sim 1$ and
take $Ma \ll 1$.
Then
\ba
\frac{1}{2a} \Omega_{0,nn}^{-1} &\approx &
\frac{1}{L} \sum_{k=1}^{N/4} 
\frac{1}{\sqrt{\frac{4\pi^2 k^2}{L^2} + M^2}} 
+\frac{1}{L} \sum_{k=N/4}^{N/2} \frac{L}{2N} \nonumber \\
&=& \frac{1}{2\pi} \int_{q_0}^{q_\infty} \frac{dq}{\sqrt{q^2+M^2}} + \frac{1}{8}
\ea
where $q \equiv 2\pi k/L$, $q_0=2\pi/L$ and $q_\infty = \pi N/(2L)$. The
integral can be evaluated to give
\ba
\frac{1}{2a} \Omega_{0,nn}^{-1} &\approx&
\frac{1}{4\pi} \ln \left [ \frac{\sqrt{p^2+M^2}+p}{\sqrt{p^2+M^2}-p} \right ]_{q_0}^{q_\infty}
+ \frac{1}{8} \nonumber \\
&\approx & \frac{1}{4\pi} \ln \left [ \frac{2}{M^2/(2q_\infty^2)} \right ] - 0 +\frac{1}{8}
\nonumber \\
&\approx& 
\frac{1}{2\pi} \ln (q_\infty a) -\frac{1}{2\pi}\ln(Ma)+ \frac{1}{4\pi} \ln (4) + \frac{1}{8} 
\nonumber \\
%&\approx&
%\frac{1}{2\pi}\ln(N) -\frac{1}{2\pi}\ln(ML) + C
%%\frac{1}{2\pi}\ln(N) -\frac{1}{2\pi}\ln(ML) + \frac{1}{2\pi}\ln(\pi) + \frac{1}{8}
%\nonumber \\
&\approx& -\frac{1}{2\pi}\ln(Ma) + C
\label{evalOm}
\ea
where we have used $M L \gg 2\pi$, in which case the $q_0$ contribution 
approximates to 0. We have also denoted the remaining terms by $C$ as
these are sensitive to the approximations we have made. 

Next we consider the consequences of changing the lattice spacing. If we
rescale $a$ to $\xi a$ for some constant $\xi$, then $\Omega_{0,nn}^{-1}/2a$ 
shifts by $-\ln(\xi)/2\pi$. This shift contributes to the mass of $\phi$ and can
be compensated for by introducing a suitable bare mass contribution in the classical
potential $V(\phi)$. Then the physical mass of $\phi$ will not depend on
rescalings of the lattice spacing. However, we still need a measurement
to tell us the physical mass of $\phi$ at a given renormalization
scale. This is normally determined by experiment.
For our purposes, we will take the renormalization scale $\mu$ to be $1/a$.
If we wish to use a different lattice spacing, say $a \to \xi a$, then
to compare results we must also change the potential: 
$V \to V + \lambda \ln(\xi) \phi^2/4\pi$.
%
%if we measure the physical mass of $\phi$ at a given energy scale, 
%we can fix the constant $a$. 
%Equivalently, the lattice spacing sets a renormalization 
%mass scale $\mu \equiv 1/a$.

The existence of the energy scale $\mu$ is also necessitated by our treatment
of $\phi$ as a classical background field. Strictly, $\phi$ should also be quantized. 
In those cases that $\phi$ can effectively be described
as a classical background, we expect the classical treatment to break
down if we probe the background on very short length scales, that is,
at very high energies. For example, if the classical
background is the spacetime metric, we expect that a quantum treatment 
will become essential at energies above the Planck scale. Similarly for
a solitonic background, we might expect that a quantum treatment will
become necessary for energy much larger than the mass scale of the 
soliton.

Now from \eqref{phicqceqstatic}, we see that the
CQC formulation for static, homogeneous $\phi_0$
is completely equivalent to the effective potential,
\ba
V_{\rm eff}(\phi_0) &=& V(\phi_0) 
+  \frac{\lambda}{2a}  \int^{\phi_0} d\phi \, \Omega_{0,nn}^{-1}[\phi ] \, \phi
\nonumber \\
&& \hskip -2 cm
\approx 
V(\phi_0) - \frac{M^2}{4\pi} \ln (Ma)
+ \lambda \left ( C+\frac{1}{4\pi} \right ) \frac{\phi_0^2}{2}
+\frac{m^2}{8\pi}
%\\
%&\approx& 
%V(\phi_0) - \frac{1}{8\pi\lambda} \left [ M^2 (2\ln (Ma) -1) \right ] 
%+ \frac{C}{2}\phi_0^2 \nonumber
\label{veff}
\ea
where $M$ is given by \eqref{Mdef}. Note that 
$-\ln(Ma) = + \ln (\mu/M) > 0$ since $\mu$ is an ultra-violet cutoff.

To summarize, we will write \eqref{phicqceqstatic} for the static, homogeneous 
background case as
\be
 \frac{\lambda}{2a} \Omega_{0,nn}^{-1} \phi_0 = - V'(\phi_0) .
 \label{phi0eq}
\ee
%where we assume that the lattice spacing $a$ corresponds to the renormalization 
%scale at which the potential is $V(\phi)$.

We now consider static solution for two simple choices for $V(\phi)$, namely
a linear potential and an inverted quadratic potential.

\subsection{Conditions for static solutions in simple cases}

\subsubsection{Linear potential}

\begin{figure}
      \includegraphics[width=0.45\textwidth,angle=0]{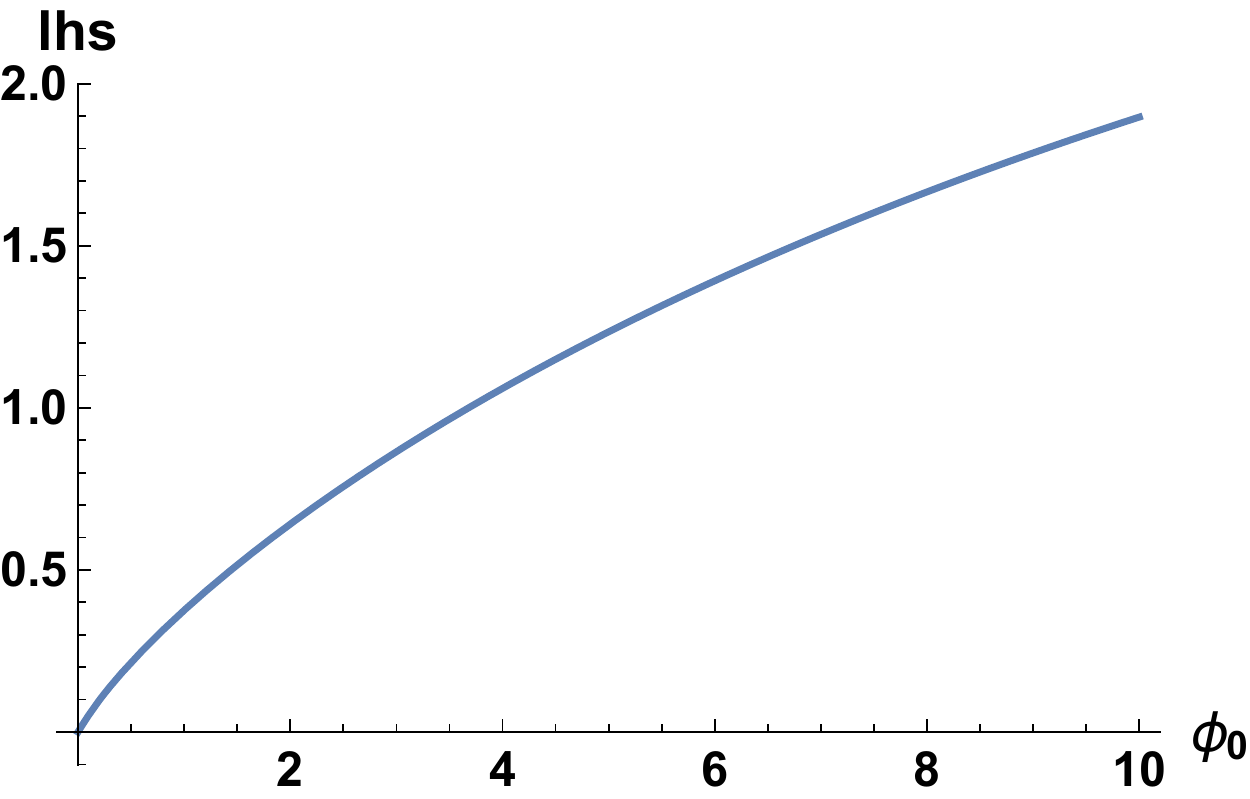}
  \caption{Plot of the left-hand side of \eqref{phi0eq} versus $\phi_0$ for 
  $\lambda=0.5$, $a=0.001$, $L=100$, $m=0$.}
  \label{plotOmegaInvTimesPhi}
\end{figure}

First we consider a linear potential
\be
V_1(\phi) = - K \phi
\label{linpot}
\ee
Then \eqref{phi0eq} becomes
\be
 \frac{\lambda}{2a} \Omega_{0,nn}^{-1} \phi_0 = +K
\ee
In Fig.~\ref{plotOmegaInvTimesPhi} we have plotted the left-hand side
of this equation. The right-hand side will be a horizontal line at $K$ and
there is clearly a solution. However, if $K$ is larger than the cutoff value 
of the left-hand side, {\it i.e.} when it is evaluated at $M=\mu$, then we cannot be
sure that there is a solution since \eqref{phi0eq} is only valid below the cutoff.
Hence the condition for a solution is
\be
0<K < K_{\rm max}
\ee
where
\be
K_{\rm max} = \sqrt{\lambda}\, \mu  \left [ \frac{1}{2a} 
\Omega_{0,nn}^{-1} \right ]_{M=\mu} \approx 
C \sqrt{\lambda}\, \mu
%\frac{\sqrt{\lambda}\, \mu}{\pi}
\ee
where we have used \eqref{evalOm}.
Since for large $\phi_0$ the potential is quadratic and increasing, the 
solution is a minimum.
Hence there is a non-trivial minimum of the potential that is entirely due to
quantum vacuum fluctuations of $\psi$ provided the coupling is strong, 
as given by 
\be
\lambda \gtrsim \frac{K^2}{C^2 \mu^2} = 2.4 a^2 K^2
%\pi^2 K^2/\mu^2
\label{critlinear}
\ee
where the value of $C$ is determined using \eqref{Omexp} with $M=\mu = 1/a$ 
and gives $C =0.643$. For weaker couplings, there is no solution for non-trivial
$\phi_0$ within the range of values in which our treatment holds.

%Eq.~\eqref{critlinear} is based on a static analysis. In Sec.~\ref{dynamics} we will
%consider dynamics on a linear potential and find some striking modifications to
%the picture. 

\subsubsection{Inverted quadratic potential}

Next we consider an inverted quadratic potential
\be
V_2(\phi)= - \frac{\kappa^2}{2} \phi^2
\ee
Then \eqref{phi0eq} becomes
\be
 \frac{\lambda}{2a} \Omega_{0,nn}^{-1} \phi_0  = +\kappa^2 \phi_0
\label{quadeq}
\ee
The left-hand side is plotted in Fig.~\ref{plotOmegaInvTimesPhi} while
the right-hand side is a straight line passing through the origin and with
slope $\kappa^2$. There is a non-trivial intersection point if the slope
$\kappa^2$ is less than the slope of the left-hand side at $\phi_0=0$
and greater than the slope of the line joining the origin to the point
where the left-hand side is evaluated at the cutoff value $\mu/\sqrt{\lambda}$.

\begin{figure}
  \includegraphics[width=0.45\textwidth,angle=0]{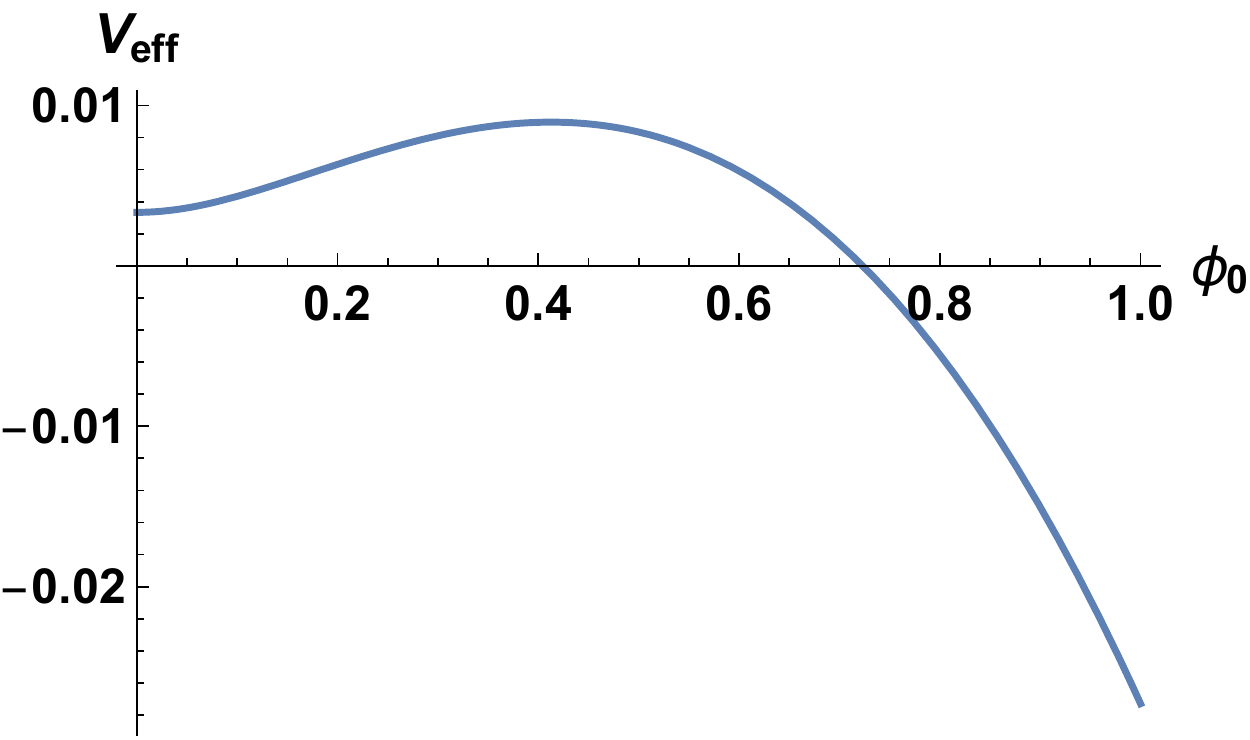}
  \caption{Plot of the effective potential for the inverted quadratic case for 
  $\kappa=1$, $m_\psi =0.1$, $a=0.25 =1/\mu$,
and $\lambda = 1.0$. As given below \eqref{critlinear}, $C=0.643$.}
  \label{quadeffpot}
\end{figure}

Near the origin, we can expand the left-hand side (lhs) of
\eqref{quadeq} for small $\phi_0$
\be
{\rm lhs}(\phi_0 \to 0) = \lambda \left [ \frac{1}{2\pi}\ln(\mu /m) + C \right ] \phi_0.
\ee
and the coefficient of $\phi_0$ is the slope at the origin. 
At the cutoff, the left-hand side evaluates to
\be
{\rm lhs}(\phi_0 = \mu/\sqrt{\lambda}) =
\lambda \left [ -\frac{m^2}{4\pi \mu^2} + C \right ] \phi_0
\ee
and the slope of the line joining the origin with the cutoff point is given
by the pre-factor of $\phi_0$.
Therefore we only have a non-trivial ($\phi_0\ne 0$) solution if
\be
 \lambda \left [ \frac{1}{2\pi}\ln(\mu /m) + C \right ] > \kappa^2 > 
 \lambda \left [ -\frac{m^2}{4\pi \mu^2} + C \right ]
\ee
which we can also write as
\be
\frac{\kappa^2}{C- m^2/4\pi \mu^2} > \lambda >
\frac{\kappa^2}{C+\ln(\mu/m) /2\pi }
\label{quadlambda}
\ee
 To understand the range of couplings for which there is a solution,
 note that $\lambda$ cannot be too small because then the quantum
 effects are negligible. On the other hand a very large value of
 $\lambda$ means that the quantum effects are very strong and
 make the classical inverted potential upright at all $\phi_0$.
 Then the only solution is the trivial $\phi_0=0$. However, evaluating
 the second derivative of the effective potential at $\phi_0=0$ shows
 that it is positive if the conditions in \eqref{quadlambda} are
 satisfied. This implies that the effective potential has a minimum
 at the origin and the non-trivial solution is a maximum.
 Thus the quantum corrections for the inverted quadratic potential
 can provide a metastable vacuum at $\phi_0=0$ in the range
 of parameters in \eqref{quadlambda} as shown in the example
 in Fig.~\ref{quadeffpot}.
 
\section{Dynamics}
\label{dynamics}

The effective potential is not suitable for describing the evolution of
the background because the derivation assumes that the quantum
field $\psi$ is in its vacuum. In the dynamical problem, as the field
$\phi$ rolls, quanta of $\psi$ are excited and the field $\psi$ is no
longer in its vacuum. The production of $\psi$ quanta backreacts 
on the dynamics of $\phi$. We shall now solve this dynamical problem, 
separately considering homogeneous and inhomogeneous backgrounds.

\subsection{Dynamics with homogeneity}
\label{withhom}

The first question we ask is if the initial conditions for $\phi$ are homogeneous, can
the dynamics make $\phi$ inhomogeneous? As this is a dynamical question, we use
the CQC equations in \eqref{zcqceq} and \eqref{phicqceq} and check that homogeneous
evolution is self-consistent.

For homogeneous $\phi$, Eq.~\eqref{Omsquared0} can be written in a more convenient 
way as
\be
\Omega^2_{ij} = -\nabla^2_{ij} + M^2 \delta_{ij}
\label{Omdelta}
\ee
where $M^2 = m^2 + \lambda \phi^2$ and the Laplacian matrix is given by,
\be
a^2 \nabla^2_{ij}= \delta_{i+1,j}-2 \delta_{ij}+\delta_{i-1,j}
\ee
where the indices are integers mod $N$. The $\nabla^2$ has translational
symmetry, {\it i.e.} 
\be
\nabla^2_{ij} = \nabla^2_{i+s,j+s}
\ee
where $s$ is any integer. Alternately, $\nabla^2_{ij}$ only depends on the
difference $i-j$ (mod $N$). Then $\Omega^2_{ij}$ is also translationally
invariant and only depends on the difference $i-j$
\be
\Omega^2_{i+s,j+s}=\Omega^2_{i,j}
\label{Om+s}
\ee
In particular, this implies
$\Omega^2_{nn}$ is independent of $n$, as already discussed below
\eqref{Omint}.

Using Eqs.~\eqref{om2}, \eqref{Osol} and \eqref{Dsol} we can check that
the initial conditions for $Z_{ij}$ are also translationally invariant,
\be
Z_{0; i+s,j+s}= Z_{0;i,j}, \ \ 
{\dot Z}_{0;i+s,j+s} = {\dot Z}_{0;i,j}
\label{Z0+s}
\ee
when $\phi$ is homogeneous. 

Next we consider the equation for $Z_{n+s,j+s}$ in \eqref{zcqceq}.
\ba
0&=&{\ddot Z}_{n+s,j+s} + \sum_{k=1}^N \Omega^2_{n+s,k} \, Z_{k,j+s} \nonumber \\
&=&
{\ddot Z}_{n+s,j+s} + \sum_{l=1}^N \Omega^2_{n+s,l+s} \, Z_{l+s,j+s} \nonumber \\
&=&
{\ddot Z}_{n+s,j+s} + \sum_{l=1}^N \Omega^2_{n,l} \, Z_{l+s,j+s}
\ea
In the above
derivation we have changed the summation index from $k$ to $l+s$ in
the second line and used \eqref{Om+s} in the third line.
Now subtracting \eqref{zcqceq} gives
\be
({\ddot Z}_{n+s,j+s} - {\ddot Z}_{n,j}) 
+ \sum_{l=1}^N \Omega^2_{n,l} ( Z_{l+s,j+s}- Z_{l,j}) =0
\ee
With the initial conditions in \eqref{Z0+s}, the solution is
\be
Z_{n+s,j+s} (t) = Z_{n,j}(t)
\ee
{\it i.e.} $Z_{n,j}$ is invariant under translations while $\phi$ is homogeneous.

Making use of the translational symmetry, we can write $Z_{nj}= a \chi_{n-j}$. Then 
going back to the equation for $\phi$ in \eqref{phicqceq}, we see
\be
\frac{1}{a^2}\sum_{j=1}^N Z_{nj}^* Z_{nj} = \sum_{j=1}^N \chi_{n-j}^* \chi_{n-j} \nonumber \\
= \sum_{k=1}^N \chi_k^* \chi_k
\label{ZZchichi}
\ee
which is independent of $n$. Thus the $\phi_n$ equation is independent
of $n$ and the evolution of $\phi$ is homogeneous.
Thus homogeneous initial conditions will lead to homogeneous evolution.

\subsection{CQC for fields with homogeneous background}
\label{cqcfieldshom}

The result above, that translational symmetry of the background is
preserved on evolution, suggests that the system of CQC equations simplify 
when the background is homogeneous. Indeed we will show here that translational
symmetry of the background implies that our quantum system corresponds
to a classical field theory. Whereas the quantum system has the real scalar
fields $\phi$ and $\psi$, the classical system has the background $\phi$
and a {\it complex} scalar field $\chi$ that is to be evolved with specific
initial conditions.

For homogeneous backgrounds we have already introduced
$\chi_{n-j}=Z_{nj}/a$ above \eqref{ZZchichi}. Then the $\phi$ equation 
becomes
%\ba
%{\ddot \phi_n} - \frac{1}{a^2} (\phi_{n+1} - 2\phi_n + \phi_{n-1}) 
%+ V'(\phi_n) && \nonumber \\
%%- \frac{\lambda}{2\pi} \ln ( \mu a ) \phi_n
%+ \lambda \left (  \sum_{j=1}^N V_j^* V_j \right ) \phi_n &=& 0 
%\label{phicqceqhom}
%\ea
\be
{\ddot \phi} + V'(\phi) + \lambda \left (  \sum_{j=1}^N \chi_j^* \chi_j \right ) \phi = 0 
\label{phicqceqhom}
\ee
where we have written $\phi_n$ as $\phi$ since it is
homogeneous. Similarly, after some manipulation, \eqref{zcqceq}
with \eqref{Omdelta} leads to
\ba
{\ddot \chi}_n - \frac{1}{a^2} (\chi_{n+1} - 2\chi_n + \chi_{n-1}) +
M^2 \chi_n = 0
\label{chieqhom}
\ea
which is the discretized version of
\be
\square \chi + M^2 \chi = 0
\ee
where $\square$ is the D'Alembertian operator and note that $\chi$ is
complex. Hence the original
system where we had a classical field $\phi$ and a quantum field
$\psi$ has been transformed into a system with $\phi$ and a
classical complex field $\chi$. 

We would now like to solve
the system of equations in \eqref{phicqceqhom} and \eqref{chieqhom}
with initial conditions following from those specified in Sec.~\ref{CQCdiscrete},
\be
\chi_q (t=0)=\frac{-i}{\sqrt{2a} N} \sum_{k=1}^N
\frac{e^{-ikq2\pi/N}}{\left [ \frac{4}{a^2}\sin^2 \left ( \frac{\pi k}{N}\right ) + M_0^2 \right ]^{1/4}}
\ee
\ba
{\dot \chi}_q (t=0)&=&\frac{1}{\sqrt{2a} N} \sum_{k=1}^N
\left [ \frac{4}{a^2}\sin^2 \left (\frac{\pi k}{N} \right ) + M_0^2 \right ]^{1/4} \nonumber \\
&& \hskip 2.5 cm 
\times e^{-ikq2\pi/N} 
\ea
where $M_0^2=m^2+\lambda \phi(t=0)^2$.

We can simplify the equations further by performing a discrete
Fourier transform,
\be
\chi_n = \frac{1}{\sqrt{N}} \sum_k c_k(t) e^{-in k2\pi/N}
\ee
Then the equation for the mode coefficients $c_k$ are ordinary differential equations
\be
{\ddot c}_k + \left [ \frac{4}{a^2}\sin^2 \left (\frac{\pi k}{N} \right ) + M^2 \right ] c_k = 0
\label{ckeq}
\ee
with the initial conditions
\be
c_k (t=0) = 
\frac{-i}{\sqrt{2aN}} 
\left [ \frac{4}{a^2}\sin^2 \left ( \frac{\pi k}{N}\right ) + M_0^2 \right ]^{-1/4}
\label{ckt0}
\ee
\be
{\dot c}_k (t=0) = \frac{1}{\sqrt{2aN}} 
\left [ \frac{4}{a^2}\sin^2 \left (\frac{\pi k}{N} \right ) + M_0^2 \right ]^{1/4} 
\label{dotckt0}
\ee
Further reduction in the number of variables can be obtained at the cost of
introducing some non-linearity by letting
\be
c_k = \rho_k e^{i\theta_k}
\ee
Then angular momentum ($\rho_k^2{\dot \theta}_k$) conservation together with the 
initial conditions gives
\be
{\dot \theta}_k = \frac{1}{2L \rho_k^2}
\ee
and the equation for $\rho_k$ is
\be
{\ddot \rho}_k + \left [ \frac{4}{a^2}\sin^2 \left (\frac{\pi k}{N} \right ) + M^2 \right ] \rho_k 
= \frac{1}{4L^2\rho_k^3}
\ee
with initial conditions
\ba
\rho_k(0)&=& \frac{1}{\sqrt{2L}} 
\left [ \frac{4}{a^2}\sin^2 \left ( \frac{\pi k}{N}\right ) + M_0^2 \right ]^{-1/4}, \\
{\dot \rho}_k(0)&=&0.
\ea
Advantage can also be taken of the symmetry $k \to N-k$ and then we only
need to solve for $N/2+1$ of the $\rho_k$'s.

In terms of the $\rho_k$'s, the equation for $\phi$ is,
\be
{\ddot \phi} + V'(\phi) + \lambda \sum_{k=1}^N \rho_k^2  \, \phi = 0 
\label{phick}
\ee

To summarize our results in this section, a quantum real scalar field in a homogeneous 
time-dependent (``rolling'') background field is equivalent to a classical complex scalar 
field in the same background with specific interactions with the background and specific
initial conditions. In the discretized version, the quantum rolling problem is thus
equivalent to $2N+1$ (recall that $c_k$'s are complex) second order ordinary 
differential equations 
\eqref{ckeq}, \eqref{phick} with the initial conditions \eqref{ckt0},
\eqref{dotckt0} and chosen initial conditions for homogeneous $\phi$. The
problem can be reduced to $(N/2+1)+1$ second order ordinary differential equations
by going to the real $\rho_k$ variables and using the $k \to N-k$ symmetry.

%The homogenous background equations are simpler than the full CQC equations. 
%However, we have chosen not to restrict our numerical analysis
%to the homogeneous background case because we also want to
%analyze inhomogeneous backgrounds.

\subsection{Dynamics in a linear potential}
\label{dynlinpot}

\begin{figure}
      \includegraphics[width=0.44\textwidth,angle=0]{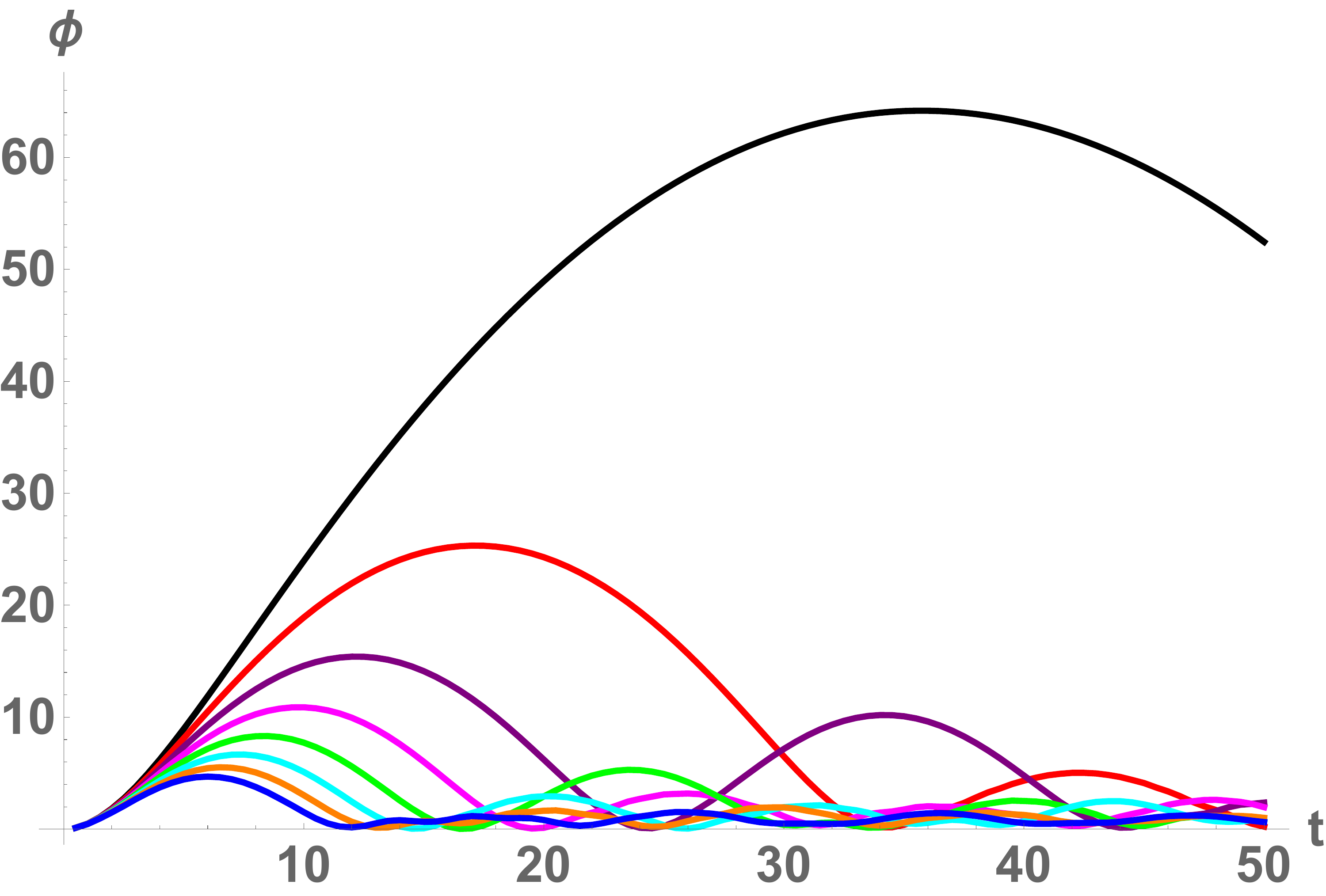}
  \caption{Plot of $\phi$ versus time 
  for $\lambda= 0.3$ (black), 0.4 (red), 0.5 (dark purple), 0.6 (light purple), 
  0.7 (dark green),
  0.8 (light green), 0.9 (orange) and 1.0 (blue), 
  and other parameters as given in the text. }
  \label{phivst}
\end{figure}

%\begin{figure}
%      \includegraphics[width=0.44\textwidth,angle=0]{phiZZlambda1}
%  \caption{Plot of $Z^*Z/a^2$ versus time for $\lambda= 0.25$ (blue), 
%  0.5 (green), 0.75 (red), and 1.0 (black).
%\TV{left plot still to be made}}
%\label{plotZZvst}
%\end{figure}

\begin{figure}
\includegraphics[width=0.44\textwidth,angle=0]{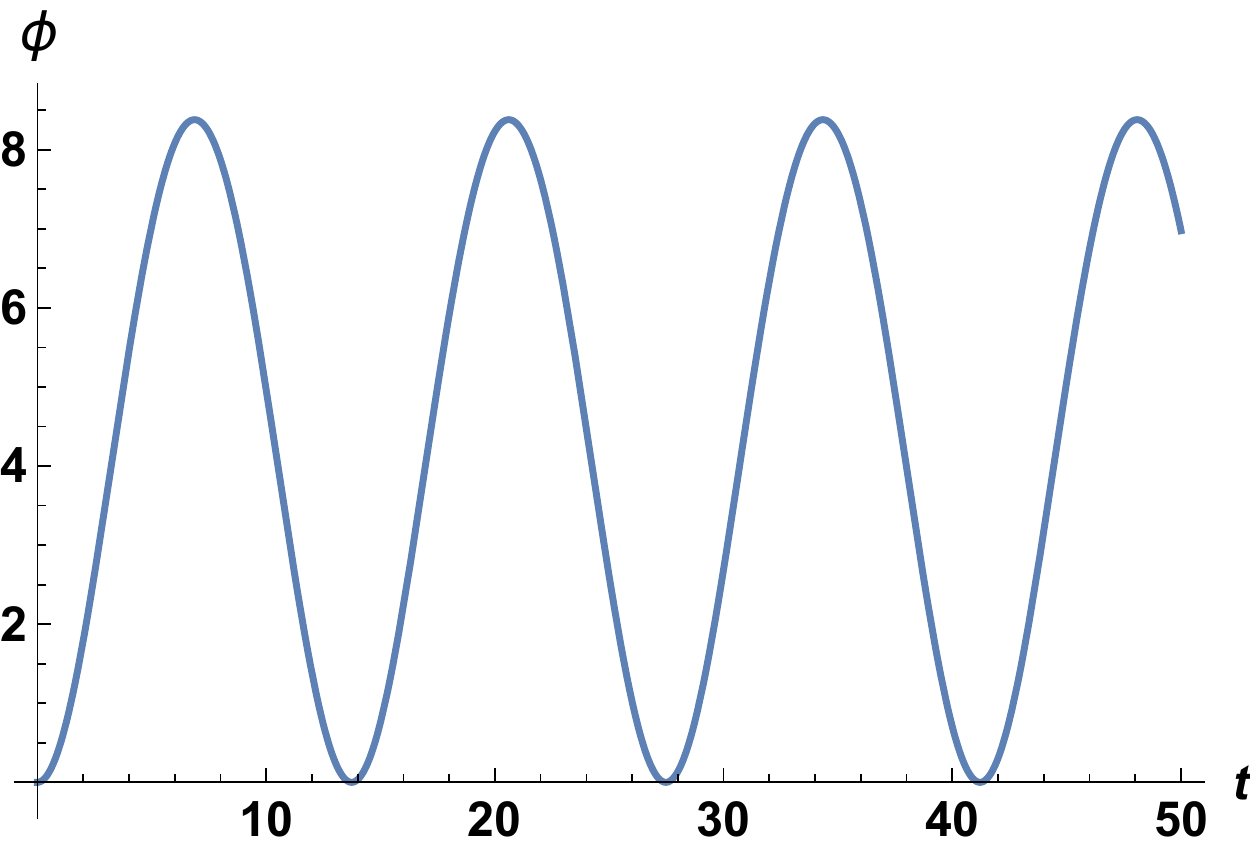}
  \caption{Plot of $\phi$ versus time for $\lambda= 0.3$ using the
  effective potential. This is to be contrasted with the CQC
  solution for $\lambda=0.3$, shown as the black curve in
  Fig.~\ref{phivst}.}
  \label{comparecqceff}
\end{figure}

For the particular case of a linear potential, we have solved for the evolution of
$\phi$ using the CQC equations in Sec.~\ref{CQCdiscrete} with the potential in
\eqref{linpot} ($K=-1$) and the $\phi$ initial conditions $\phi_n(0)=0={\dot \phi}_n(0)$.
The solutions for $\phi (t)$ for several different values of $\lambda$ and
 with parameters $a=0.25$, $N=400$, $L=100$, $m=0.1$,
are shown in Fig.~\ref{phivst}. 

The plots show that $\phi$ does not
increase monotonically as we might expect based on rolling on a
classical linear potential; instead $\phi$ oscillates, as one might expect 
based on the effective potential analysis. To compare the CQC dynamics
with that of rolling on the effective potential, we have solved the 
``effective equation of motion'',
\be
{\ddot \phi} + V_{\rm eff}'(\phi) = 0
\ee
where $V_{\rm eff}(\phi)$ is given in \eqref{veff} and $V(\phi)=-\phi$.
Fig.~\ref{comparecqceff} shows the rolling solution on the effective
potential for $\lambda=0.3$. It is to be compared to the corresponding
curve in Fig.~\ref{phivst}.

\begin{figure}
\includegraphics[width=0.44\textwidth,angle=0]{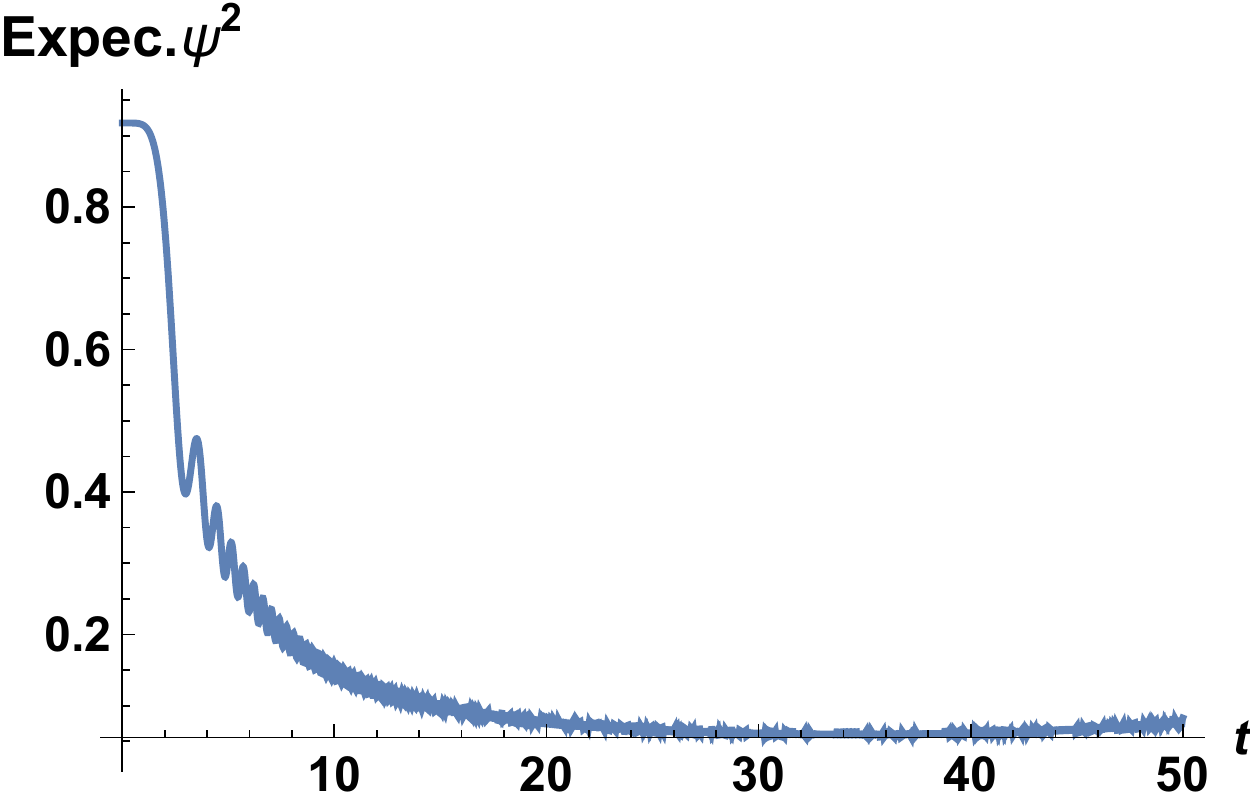}
  \caption{Plot of $\langle \psi^2 \rangle$ versus time for $\lambda= 0.3$.}
  \label{expecpsi2}
\end{figure}

A few features of the dynamics stand out: the field $\phi$ oscillates in the 
full dynamics (Fig.~\ref{phivst}) but at a much smaller frequency than in 
the effective potential treatment (Fig.~\ref{comparecqceff});
the amplitude of oscillations in the effective potential stays constant and 
is much smaller than in the CQC. This is surprising since the physical
argument is that $\psi$ particles are produced during rolling and this
is what causes differences between the full dynamics and the dynamics
on the effective potential. However, increased particle production might
be expected to increase $\langle \psi^2\rangle$ and this should cause 
the $\phi$ oscillations in the full dynamics to have {\it smaller} amplitude 
than in the effective potential. The resolution is that even though there
is particle production, $\langle \psi^2\rangle$ actually {\it decreases} as is
evident in Fig.~\ref{expecpsi2}. This can happen if most of the energy
in particle production goes into the kinetic energy and not in 
$\langle \psi^2\rangle$. Then with a smaller $\langle \psi^2\rangle$
we do expect the $\phi$
oscillations to have larger amplitude in the full dynamics.

\begin{figure}
      \includegraphics[width=0.44\textwidth,angle=0]{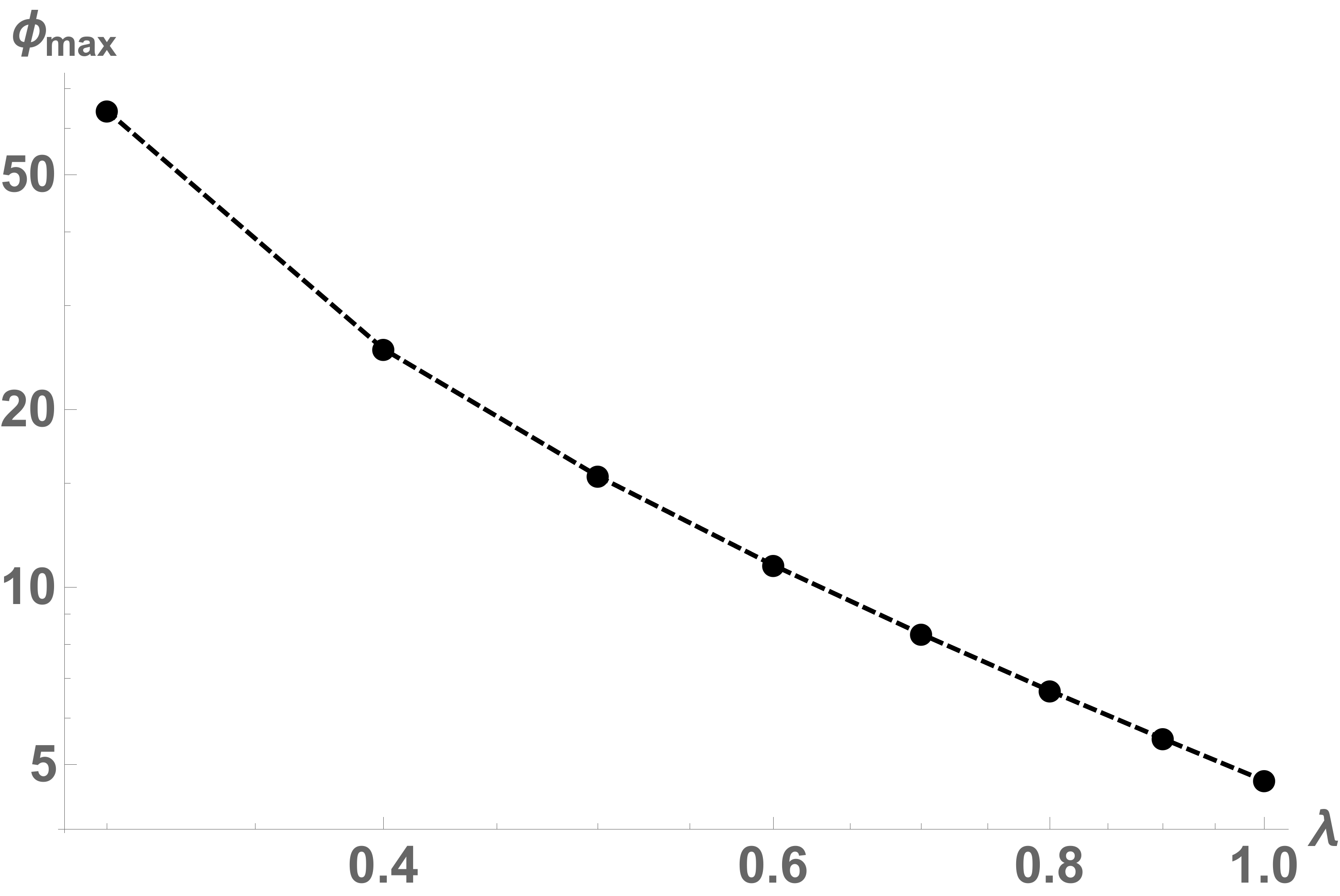}
  \caption{Log-log plot of the value of $\phi$ at the first turning point versus 
  $\lambda$. The fit is not a straight line, ranging from $\phi_{\rm max} \sim \lambda^{-3.2}$
  for smaller values of $\lambda$ and $\sim \lambda^{-1.7}$ for the larger values.
   }
\label{phiturnvslambda}
\end{figure}

\begin{figure}
      \includegraphics[width=0.44\textwidth,angle=0]{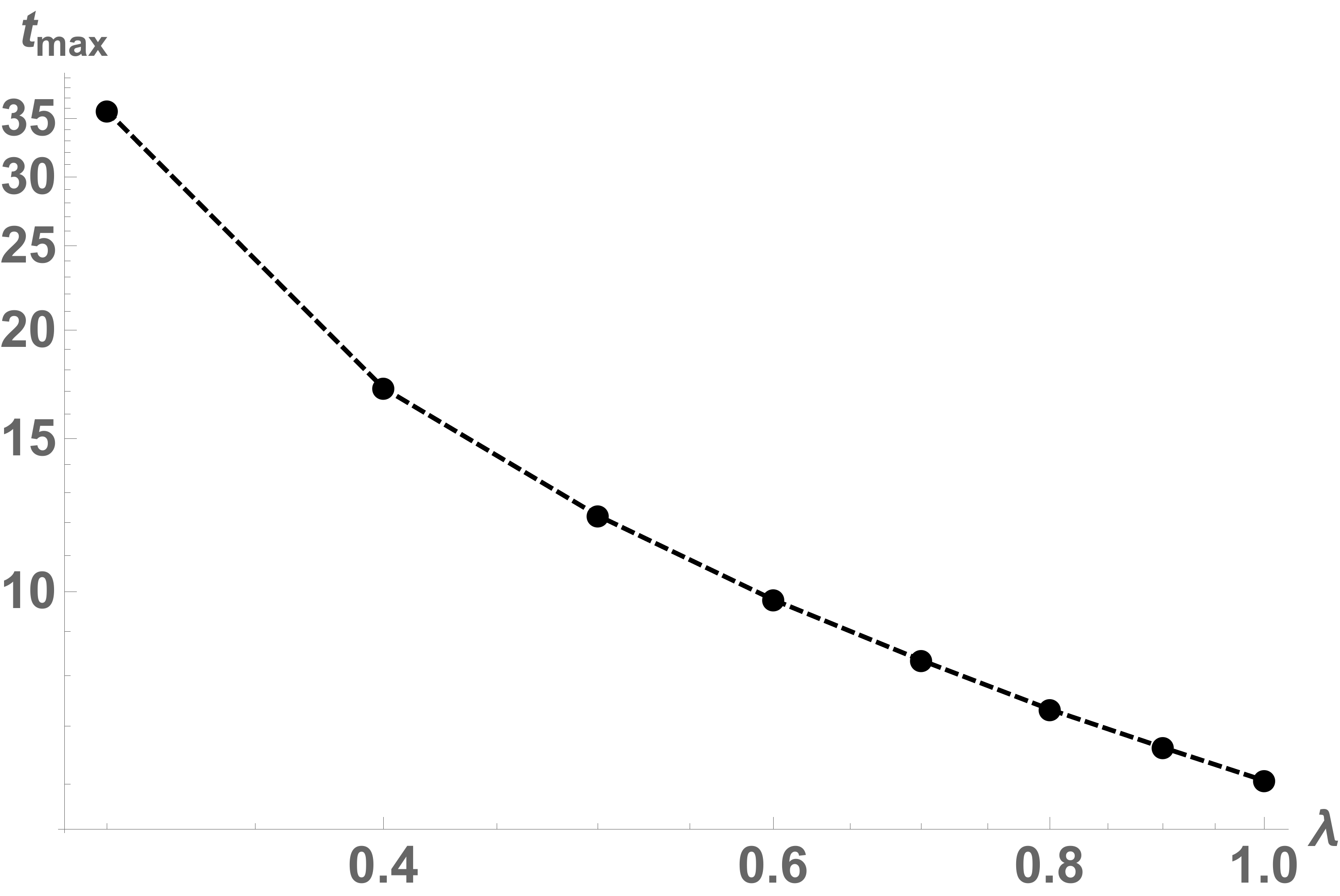}
  \caption{Log-log plot of the value of $t$ at the first turning point versus 
  $\lambda$.  The fit is not a straight line, ranging from $t_{\rm max} \sim \lambda^{-1.9}$
  for smaller values of $\lambda$ and $\sim \lambda^{-0.9}$ for the larger values.
  }
\label{tturnvslambda}
\end{figure}

In the CQC solution, let us denote the first maximum value
of $\phi$ by $\phi_{\rm max}$ and the time at which this value is
reached by $t_{\rm max}$. In Fig.~\ref{phiturnvslambda} we show
$\phi_{\rm max}$ as a function of $\lambda$ on a log-log plot. It is
clear that the data is not fit by a power law as the fit varies from
$\phi_{\rm max} \sim \lambda^{-3.2}$ for smaller $\lambda$ to 
$\phi_{\rm max} \sim \lambda^{-1.7}$ at larger $\lambda$.
Fig.~\ref{tturnvslambda} shows $t_{\rm max}$ versus $\lambda$ on
a log-log plot. Here too the fit varies from $t_{\rm max} \sim \lambda^{-1.9}$
to $\sim \lambda^{-0.9}$ at larger $\lambda$.

%We can understand the oscillatory behavior of $\phi$ for all
%(non-zeo) values of the interaction strength $\lambda$ as due
%to particle production. At $t=0$ the $\phi$ field starts rolling down the linear 
%potential, and as it continues to roll, more and more quanta of $\psi$ are excited. 
%The backreaction of the quanta on the dynamics of $\phi$ depends on the
%expectation value of $\psi^2$ in the instantaneous quantum state of $\psi$.
%%(In contrast, the effective potential is given by the expectation value evaluated
%%in the quantum ground state.) 
%The expectation value will depend on the
%accummulated particles, that will be related to the amount that $\phi$
%has been displaced and also to the coupling constant $\lambda$. Thus
%the backreaction term in the CQC equation of motion for $\phi$ 
%will go like $\lambda^2 \phi^2$, and will dominate the linear potential
%term for large $\phi$. This rough argument supports the observed oscillatory 
%behavior though cannot be expected to be quantitatively accurate.

Even though we have shown that homogeneous initial conditions lead
to homogeneous evolution, there remains the possibility that the evolution 
is unstable to developing inhomogeneities. We now address this question 
numerically by including small perturbations to homogeneous initial conditions.

\subsection{Dynamics with small initial inhomogeneities}
\label{inhomdyn}

To introduce inhomogeneous perturbations, we solve the CQC equations
in \eqref{phicqceq}, \eqref{zcqceq} but with the initial conditions
\be
\phi_n(t=0)=0, \ \  {\dot \phi}_n (t=0) = \epsilon \sin\left ( \frac{2\pi n \nu}{N} \right )
\ee
where $\epsilon$ is a small amplitude and the integer $\nu$ sets the
wavenumber of the perturbation.
The $Z$ initial conditions are still given by \eqref{Zinitial}. The
advantage of introducing the inhomogeneities in the time derivative
${\dot \phi}_n$ while
keeping $\phi_n$ homogeneous is that then we can continue to use
\eqref{om2} with the formula for $O$ and $D$ given in Sec.~\ref{hsoln}.

\begin{figure}
      \includegraphics[width=0.44\textwidth,angle=0]{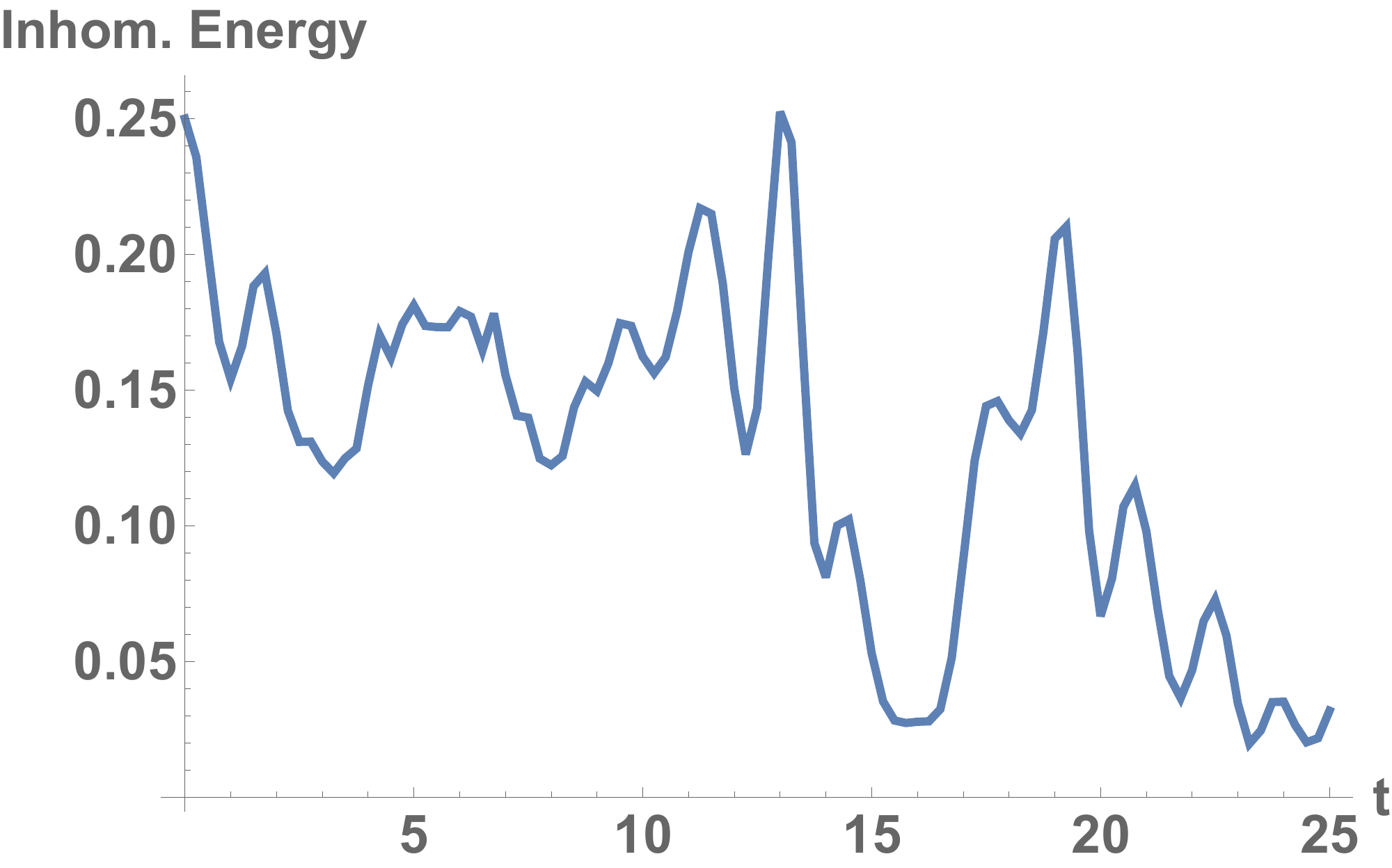}
  \caption{Energy in inhomogeneity versus time for $\lambda = 1.0$
  $\epsilon=0.1$ and $\nu=N/10$.
 }
\label{inhomEvst}
\end{figure}

We now write the field $\phi$ as
\be
\phi = {\bar \phi} + {\delta \phi}
\ee
where the homogeneous part is
\be
{\bar \phi}(t) \equiv \frac{1}{N} \sum_{n=1}^N \phi_n
\ee
The energy in the inhomogeneous part is
\be
E_{\rm inhom} = a \sum_{n=1}^N \left [ \half ( {\delta {\dot \phi}}_n)^2
+ \frac{1}{2} \left ( \frac{{\delta \phi}_{n+1}-{\delta \phi}_{n-1}}{2a} \right )^2
\right ]
\ee
In Fig.~\ref{inhomEvst} we show $E_{\rm inhom}$ versus $t$ for
$\lambda=1.0$, $\epsilon=0.1$ and $\nu = N/10$. It is clear that the
energy in the inhomogeneities decreases with time, though with some
fluctuations, and there is no instability in the system. We find similar
evolution for other values of $\nu$.

\section{Conclusions}
\label{conclusions}

We have solved for the dynamics of a classical rolling field that is coupled to a
quantum field using the CQC. 

Static solutions of the CQC
equations are simply the extrema of the effective potential. For the
particular model in \eqref{model} with the linear potential of
\eqref{linpot}, the effective potential has a minimum in the regime of
validity of our equations only if the
interaction strength is stronger than a critical value as in \eqref{critlinear}.
For weaker interactions, there may be a minimum but it would lie
beyond our cutoff.

The CQC equations are then used to study the dynamics of rolling on the 
linear potential. With homogeneous initial conditions, we find that the background 
field oscillates. This is similar to what we would expect from the effective
potential picture but there are sharp differences in the details. These are
most easily seen in Figs.~\ref{phivst} and \ref{comparecqceff}
and are understood by noting that the CQC solves for the full dynamics,
including particle production and backreaction, whereas the effective
potential picture is limited to static backgrounds.

In order to study possible dynamical instabilities, we have also
examined the case when the background field is weakly inhomogeneous.
Our numerical results show that
small inhomogeneities in the initial conditions diminish on evolution and
there is no indication of an instability.

Our analysis is directly relevant to phase transitions in which the order
parameter acquires a vacuum expectation value. The CQC equations can
be used to study the dynamics of phase transitions, in particular the formation
of topological defects. However, it would become necessary to generalize the 
CQC to the case when the quantum field has self-interactions.
One way to deal with self-interactions, {\it e.g.} a $\psi^4$ term in the action, 
is to use perturbation theory on top of the CQC solution. That is, the solution
to the CQC equations would serve as the zeroth order solution around which
self-interactions could be treated perturbatively. This scheme has not yet
been implemented.

Our result that homogeneous initial conditions evolve homogeneously
is equivalent to saying that the quantum dynamics does not spontaneously
break translational invariance. This is in contrast to the claim that 
cosmological inflation due to a rolling homogeneous field
produces density fluctuations and thus spontaneously breaks translational 
symmetry~\cite{Mukhanov:1981xt}. However, further investigation of this issue 
is necessary because 
there are additional ingredients that go into the inflationary calculation. In 
particular, quantum fluctuations convert into classical fluctuations once they 
exit the cosmological horizon~\cite{Kiefer:1998qe}, and cosmological expansion 
provides dissipation. It will be interesting to capture these effects in the CQC 
formulation.

\acknowledgements
We are especially grateful to George Zahariade for important feedback, and to
Matt Baumgart, Juan Maldacena and Vincent Vennin for discussions and comments. 
TV thanks CEICO (Prague),  APC (Universite Paris Diderot), and IAS (Princeton) 
for hospitality while this work was being done. TV is supported by the U.S. Department of Energy, 
Office of High Energy Physics, under Award No.~DE-SC0019470
at Arizona State University.

\appendix

\bibstyle{aps}
\bibliography{paper}

\end{document}